# High-Fidelity Transfer of 2D Bismuth Oxyselenide and its Mechanical Properties

Wenjun Chen,[1, #] Usman Khan,[1, #] Simin Feng,[1] Baofu Ding,[1] Xiaomin Xu,[1] Bilu Liu[1, *]

[1]*Shenzhen Geim Graphene Center, Tsinghua-Berkeley Shenzhen Institute and Tsinghua Shenzhen International Graduate School, Tsinghua University, Shenzhen, 518055, P. R. China*

[#]These authors contributed equally to this work

*To whom correspondence should be addressed. E-mail: bilu.liu@sz.tsinghua.edu.cn

# ABSTRACT

Two-dimensional (2D) bismuth oxyselenide ($Bi_2O_2Se$) with high electron mobility is advantageous in future high-performance and flexible electronic and optoelectronic devices. However, transfer of thin $Bi_2O_2Se$ flakes is rather challenging, restricting measurements of its mechanical properties and application exploration in flexible devices. Here, we develop a reliable and effective polydimethylsiloxane (PDMS)-mediated method that allows transferring thin $Bi_2O_2Se$ flakes from grown substrates onto target substrates like micro-electro-mechanical system substrates. The high fidelity of the transferred thin flakes stems from the high adhesive energy and flexibility of PDMS film. For the first time, the mechanical properties of 2D $Bi_2O_2Se$ are experimentally acquired with nanoindentation method. We found that few-layer $Bi_2O_2Se$ exhibits a large intrinsic stiffness of 18-23 GPa among 2D semiconductors, and a Young's modulus of 88.7±14.4 GPa which is consistent with the theoretical values. Furthermore, few-layer $Bi_2O_2Se$ can withstand a high radial strain of more than 3%, demonstrating excellent flexibility. The development of the reliable transfer method and documentation of mechanical properties of 2D $Bi_2O_2Se$ jointly fill the gap between theoretical prediction and experimental verification of mechanical properties of this emerging material, and will promote flexible electronics and optoelectronics based on 2D $Bi_2O_2Se$.

# 1. Introduction

Two-dimensional (2D) materials have drawn significant attention for flexible electronic and optoelectronic devices.[1-3] Bismuth oxyselenide ($Bi_2O_2Se$), a semiconducting 2D material featured by narrow bandgap (~ 0.8 eV), high electronic mobility (20,000 $cm^2$ $V^{-1}$ $s^{-1}$ at 2K) and good air stability, shows great potential for applications in high-performance electronics, optoelectronics, and flexible devices.[4-10] For example, the photodetectors based on $Bi_2O_2Se$ with ultra-broadband responses,[7] high on/off ratio and ultrahigh photodetectivity[8] have been demonstrated. $Bi_2O_2Se$ is also an important material to fabricate three-terminal memristors with high speed and low energy consumption for neuromorphic functions.[10] In addition, the combination of high performance and flexibility in 2D $Bi_2O_2Se$-based optoelectronic devices has drawn much attention. Attempts have been made to directly fabricate bendable $Bi_2O_2Se$-based photodetector arrays on mica substrate with a certain extent of flexibility, which exhibits ultrafast photoresponse of ~ 1 ps and operates steadily under the bending strain of 1%.[11] Further efforts have been put forth to improve the pliability of the flexible optoelectronics based on 2D $Bi_2O_2Se$. For instance, the related photodetectors fabricated on flexible polyethylene terephthalate shows stable photocurrent after 500 bending cycles.[12] It is suggested that 2D $Bi_2O_2Se$ is of great importance in making flexible and integrated electronics for realistic applications.

Revealing the mechanical properties of 2D materials is the first prerequisite for their applications in flexible and stretchable electronics.[13-15] Indeed, the mechanical characteristics of monolayer and few-layer 2D materials, such as metallic graphene[16-19] and transition metal carbides or/and nitrides (MXenes),[20] insulating hexagonal boron nitride,[18] and semiconducting transition metal dichalcogenides (TMDCs)[21-24] and black phosphorus (BP),[25] have been intensively studied and documented by atomic force microscopy (AFM) nanoindentation. The above-mentioned 2D materials are bonded by interlayered van der Waals force, while the relatively strong electrostatic force dominates the interaction between $[Bi_2O_2]^{2n+}$ and $[Se]^{2n-}$ layers in $Bi_2O_2Se$,[4-5] making the intrinsic mechanical properties of $Bi_2O_2Se$ intriguing, yet unknown experimentally. Theoretical calculations have indicated that $Bi_2O_2Se$ possesses a high theoretical Poisson's ratio, and Young's modulus is 73.86 and 83.4 GPa calculated by Perdew-Burke-Ernzerhof (PBE) exchange-correlation functional[26] and projector augmented wave (PAW) method,[27] respectively. Unfortunately, there is so far no experimental report of mechanical properties of 2D $Bi_2O_2Se$. One key obstacle in measuring mechanical properties of $Bi_2O_2Se$ is the difficulties in transferring $Bi_2O_2Se$ flakes, particularly thin ones, from mica onto the desired substrates, as the adhesion between the $Bi_2O_2Se$ and mica substrate is very strong. Although few articles have reported that $Bi_2O_2Se$ flakes could be transferred onto other substrates, such as rigid $SiO_2/Si$[8, 28] and bendable polymer,[12] with the assistance of polymethyl methacrylate (PMMA) combined with hydrofluoric acid (HF) etching, the complicated processes and high corrosion of HF make the transfer difficult to handle, even etch the $Bi_2O_2Se$ flakes. Moreover, transfer of thin $Bi_2O_2Se$ flakes are not practical yet, which hinders measurements of their mechanical properties and further application explorations. Therefore, it is crucial to develop a method to transfer thin layers of $Bi_2O_2Se$ from mica onto target substrates to measure its mechanical features.

In this work, we develop a polydimethylsiloxane (PDMS)-mediated transfer method to successfully transfer thin $Bi_2O_2Se$ flakes onto target substrates. In this PMDS-mediate transfer method, we use PDMS to directly peel off thin $Bi_2O_2Se$ flakes from mica substrate and transfer them onto micro-electro-mechanical system (MEMS) SiO$_2$/Si substrates with periodic arrays of holes. The suspended few-layer Bi$_2$O$_2$Se flakes with high fidelity and variable thicknesses, including thin ones down to 3.2 nm (5 layers, 5L), are obtained by such PMDS-mediated transfer method, because PDMS as a laminate with high adhesive force and flexibility makes the Bi$_2$O$_2$Se flakes adhere tightly to detach from mica to fulfill the transfer. AFM nanoindentation measurements show that the stiffness of few-layer Bi$_2$O$_2$Se flakes is in the range of 18-23 GPa, which compares favorably among semiconducting 2D materials. Meanwhile, Young's modulus of Bi$_2$O$_2$Se is measured to be 88.7±14.4 GPa, in accordance with the theoretical values and is found to be layer-independent. We further reveal that the thin Bi$_2$O$_2$Se flakes maintain good flexibility and elasticity within a high radial strain up to 3%. Our work discloses the first experimental results about mechanical properties of emerging 2D Bi$_2$O$_2$Se and paves a way to explore its applications in flexible electronics and mechanical property related devices.

## 2. Results and Discussion

Few-layer and single-crystalline Bi$_2$O$_2$Se flakes on mica substrate were synthesized with vapor-solid (VS) deposition following our previous work (please see Experimental Section for details).[8] Transfer of Bi$_2$O$_2$Se flakes is essential for subsequent measurements of physical properties and further applications. The schematic illustration of the proposed PDMS-mediated and the previous established PMMA-assisted transfer[8, 29] processes are shown in **Figure 1**. For the PMMA-assisted method (Figure 1a), PMMA as a sacrificial layer was initially spin-coated on Bi$_2$O$_2$Se/mica, followed by the drying process. A piece of PDMS was attached onto PMMA/Bi$_2$O$_2$Se/mica, heated to enhance the adhesion, and then PDMS/PMMA/Bi$_2$O$_2$Se was peeled off from mica and transferred to the MEMS SiO$_2$/Si substrate. Suspended Bi$_2$O$_2$Se flakes were obtained after peeling off PDMS and immersing the sample in acetone overnight to dissolve PMMA (please see Experimental Section for details). Notably, this method is only effective in detaching thick $Bi_2O_2Se$ flakes from mica, as PMMA prohibits the direct contact between $Bi_2O_2Se$ and PDMS.

However, the PMMA-assisted method is unable to transfer thin $Bi_2O_2Se$ flakes, which remained on mica. Because rigid PMMA with low adhesive force[30] may hinder its conformable contact with $Bi_2O_2Se$ flakes that the strong interaction between the thin flakes and mica cannot be overcome. According to the measurement results by AFM contact mode, which has been used to evaluate the adhesion of polymers,[30] the adhesive force of PDMS is more than twice as high as that of PMMA (Figure S1). Therefore, the PDMS-mediated transfer method (Figure 1b), which is independent on the sacrificial layer of PMMA, is developed here to transfer $Bi_2O_2Se$ flakes. To start with, a piece of flexible PDMS with the thickness of 200 μm was directly pressed onto the as-grown sample to make a good and direct adhesion between the polymer and $Bi_2O_2Se$ flakes. Afterwards, $Bi_2O_2Se$ flakes along with the PDMS layer were peeled off from the growth substrate with the help of a water droplet of 5 μL and pasted onto the MEMS $SiO_2$/Si substrate with periodic arrays of holes. Subsequently, the sample was gradually heated from the room temperature to 180 °C to strengthen the interaction between $Bi_2O_2Se$ flakes and the target substrate. Eventually, instead of the long dissolution in solvent, the polymer layer was automatically detached and easily removed at 180 °C, leaving the thin $Bi_2O_2Se$ flakes on the MEMS $SiO_2$/Si substrate. With this method, $Bi_2O_2Se$ flakes can be transferred onto other desired substrates (Figure S2). Hence, in comparison to the PMMA-assisted processes, the PDMS-mediated transfer method offers a simpler and more time-saving pathway to accomplish the transfer of $Bi_2O_2Se$ flakes. More importantly, it is more convenient and effective to transfer thinner $Bi_2O_2Se$ flakes, which takes advantage of the high adhesive force as well as the flexibility of PDMS to make a strong adhesion with the thin flakes.

After transfer onto the MEMS $SiO_2$/Si substrate via the PDMS-mediated method, all $Bi_2O_2Se$ flakes keep the original rectangular shape and the structural integrity, as shown with the scanning electron microscope (SEM) image in Figure 1c. The pre-patterned holes on the substrate are completely or partially covered by suspended $Bi_2O_2Se$ flakes, with their shape visible from the SEM image (Figure 1c), and the fine structure presented by a high-magnification SEM image (Figure 1d). On the contrary, $Bi_2O_2Se$ flakes transferred by the PMMA-assisted method are thick, that the holes underneath are unrecognizable from the SEM image (Figure 1e). The collection of $A_{1g}$ mode (~ 159 cm$^{-1}$) in Raman spectra (Figure 1f) reveals the consistent quality of few-layer $Bi_2O_2Se$ flake before and after the PDMS-mediated transfer processes. To identify the layer number of $Bi_2O_2Se$, a simple method is developed by measuring the intensity ratio of $A_{1g}$ mode of $Bi_2O_2Se$ to the first-order transverse-optical-phonon mode of Si at ~ 520.7 cm$^{-1}$, which shows monotonically increasing dependence with the thickness of $Bi_2O_2Se$ (Figure S3). In addition, Raman mapping were performed to verify the uniformity of $Bi_2O_2Se$ flakes after the transfer process. As an example, Figure 1g shows an optical image of an as-grown $Bi_2O_2Se$ flake on the mica substrate, and the corresponding Raman map with the intensity of $A_{1g}$ peak (Figure 1h) indicates its high homogeneity. After transferring onto the MEMS $SiO_2$/Si substrate, the shape of few-layer $Bi_2O_2Se$ flakes remains intact and the expected suspended parts can be obtained, as shown in Figure 1i. The homogeneous Raman mapping in Figure 1j reflects the well-maintained uniformity of the transferred $Bi_2O_2Se$ flakes, confirming the fidelity and reliability of the PMMA-free transfer method. It is also noticeable in Figure 1i that the suspended part of $Bi_2O_2Se$ flake is distinguishable with lower peak intensity compared to that of the supported part on $SiO_2$/Si substrate.

The crystal structure of the transferred $Bi_2O_2Se$ samples was further investigated by transmission electron microscopy (TEM) characterization. The high-resolution TEM (HRTEM) image in **Figure 2**a indicates that the d-spacing of few-layer $Bi_2O_2Se$ is 0.28 nm for the (110) plane, depicting the high quality of the samples after the transfer. The ($1\bar{1}0$), (110), and (200) spots in the selected area electron diffraction (SAED) pattern (inset in Figure 2a) comply with square arrangement, indicating the single-crystalline nature of the $Bi_2O_2Se$ flakes. XPS analyses have also implemented to confirm the high quality and chemical composition of $Bi_2O_2Se$ flakes (Figure S4). The morphology and thickness of the transferred $Bi_2O_2Se$ flakes on MEMS $SiO_2$/Si substrate were further characterized by atomic force microscopy (AFM). Prior to the mechanical property tests, height profile of AFM image firstly helps locate the suspended $Bi_2O_2Se$ part covering on holes, with the depth of approximately 300 nm (Figure S5). For instance, as shown by the green line in the AFM image (Figure 2b), the corresponding height analysis in Figure 2c verifies the well-like morphology of the suspended few-layer $Bi_2O_2Se$ with the depth displacement at the center position of 65.5 nm, which is taken as a base value for the subsequent calculation of the radial strain. It is worth noting that the spontaneous deformation of the suspended flake is ascribed to its gravity. Meanwhile, the stair-stepping curve at the edge of the suspended $Bi_2O_2Se$ flake in AFM image indicates its thickness is 6.2 nm (Figure 2d and e), corresponding to 10L, which is requisite for subsequent calculations of its mechanical properties. These results suggest that the PDMS-mediate method can transfer thin $Bi_2O_2Se$ flakes with high fidelity and negligible degradation.

We use AFM nanoindentation to measure the intrinsic mechanical properties of 2D materials.[13] **Figure 3**a shows the schematic of the nanoindentation experiment, prior to which the spring constant of the AFM cantilever was measured. After the topography scanning of $Bi_2O_2Se$ which allows localization of its suspended parts, the tip was then placed at the hole center. Before applying a vertical force via the AFM cantilever, the suspended flake was pre-strained because of its gravity. The displacement of the center position was defined as $\delta_0$, which is marked in the schematics of measurement mechanism (top in Figure 3b). Subsequently, the cantilever was controlled to move downwards with a certain velocity to deflect the $Bi_2O_2Se$ flake with the indentation of $\delta$ at the same position (bottom in Figure 3b), until the breakdown point. The starting horizontal part of the initially collected force curve (Figure S6a) reflects the approaching process of the AFM tip before contacting the surface of the sample. Similar to previous reports, the crossing of the extended line of the horizontal part and the force curve was set as the original deflection position where both the force and the indentation are zero.[20, 22]

Several loading and unloading cycles were implemented at the same position for each flake to study the elastic responses during the process. Figure 3c plots the force curves of a 5L $Bi_2O_2Se$ flake under different maximum loads. All the three curves show the same tendency in the overlapping region, excluding the relative sliding between the flake and the substrate during the extension-retraction cycles, and indicating high elasticity of the $Bi_2O_2Se$ flake. In addition, the 5L flake was fractured at the applied force of 1452.5 nN with an indentation of 152.8 nm, according to the blue curve. The fracture load ($F_f$) and the maximum indentation are two key parameters utilized to calculate stiffness and flexibility, respectively. Height profiles of the suspended $Bi_2O_2Se$ (insets of Figure 3c, see the height distributions of red and blue profiles in Figure S6b) further show its well-like morphology before applying force (inset of Figure 3c, the left panel) and when the fracture is induced by overload (inset of Figure 3c, the right panel). Force curves of few-layer $Bi_2O_2Se$ with variable layer numbers (5-40L) were collected and plotted in Figure 3d, so as to comprehensively analyze the thickness-dependence of mechanical properties.

As the hole is circular, the strain applied on the flake via the tip is reasonably isotropic at the hole center. Notably, the radius of the tip ($r_t$, ~ 7 nm) is much less than that of the hole ($r$, ~ 1 μm) with a ratio of $r_t/r = 0.007 \ll 1$. Therefore, the isotropy in strain and the negligible ratio permit the use of following $F$-$\delta$ equation to fit force curves,[16, 20-21]

$$F = \sigma_0^{2D}\pi\delta + E^{2D}\frac{q^3\delta^3}{r^2} \qquad (1)$$

where $\sigma_0^{2D}$ denotes the pre-strain of the suspended flake, $E^{2D}$ is 2D elastic modulus, $q$ is a constant and expressed by $q = (1.049 - 0.15v - 0.16v^2)^{-1}$, and $v$ represents the Poisson's ratio of $Bi_2O_2Se$ with a theoretical value of 0.35.[26] It is worth mentioning that a linear relevance between $F$ and $\delta$ (first term of Eq. (1)) is found at the small loading range, while an $F$-$\delta^3$ relationship (second term of Eq. (1)) dominates at the range of large loads, as elucidated by the force curve of 6L $Bi_2O_2Se$ in logarithmic scale coordinates (inset of Figure 3d). A similar phenomenon also occurred in the measurement of mechanical properties of $Bi_2O_2Se$ with different layer numbers (Figure S7) and other 2D materials.[16, 20-21, 23] According to Eq. (1), all the fitting curves (red curves in Figure 3d) exhibit good agreement with the experimental plots ($R^2 > 0.998$). As observed, thicker suspended $Bi_2O_2Se$ flakes are capable to withstand higher mechanical force, corresponding to a higher fracture load. We have collected force curves of 28 suspended $Bi_2O_2Se$ flakes with different layer numbers from 5L to 40L. We found that upon increasing the layer number, $E^{2D}$ monotonically increases from ~ 300 to ~ 2500 N m$^{-1}$ (Figure 3e). The calculated results of $\sigma_0^{2D}$ help evaluate the pre-tension of few-layer Bi$_2$O$_2$Se flakes with different thickness, which are in the range from 0.5 to 3.5 N m$^{-1}$ (Figure S8). Afterwards, according to the combination of calculated $E^{2D}$ and measured $F_f$, the breaking strength ($\sigma_{max}^{2D}$) of few-layer Bi$_2$O$_2$Se can be obtained by,[16, 20-21]

$$\sigma_{max}^{2D} = \left(\frac{F_f E^{2D}}{4\pi r_t}\right)^{\frac{1}{2}} \quad (2)$$

it shows the same trend with $E^{2D}$ versus layer number that increases from 60 to 550 N m$^{-1}$ in the same range (Figure 3f). In general, the elastic responses of few-layer Bi$_2$O$_2$Se attest that $E^{2D}$ and $\sigma_{max}^{2D}$ are positively dependent on its thickness.

Based on the calculated $\sigma_{max}^{2D}$ and $E_{2D}$, two fundamental parameters that describe mechanical properties of the materials, i.e., stiffness ($\sigma$) and Young's modulus ($E_{Young}$), can be determined by,[16, 20-21]

$$\sigma = \frac{\sigma_{max}^{2D}}{t} \quad (3)$$

$$E_{Young} = \frac{E_{2D}}{t} \quad (4)$$

the calculated $\sigma$ of Bi$_2$O$_2$Se with different layer number from 5L to 22L and the comparison among various 2D materials are summarized in **Figure 4**a. Note that $\sigma$ of few-layer Bi$_2$O$_2$Se (18-23 GPa) is approximately one-fifth that of atomically thin 2D materials such as graphene and BN,[16-18] it is high among 2D materials with semiconducting behaviors.[21, 25] Specifically, $\sigma$ of few-layer Bi$_2$O$_2$Se flakes is comparable to that of TMDCs,[21] and four times higher than that of few-layer BP,[25] which indicates the capability of Bi$_2$O$_2$Se flakes to withstand out-of-plane external forces. Considering $E_{Young}$ of few-layer Bi$_2$O$_2$Se at the same range of layer number, which is independent on the thickness and the values are located at the range from 80-100 GPa (Figure 4b). Similar with the theoretically predicted value of 73.86 and 83.4 GPa by PBE functional[26] and PAW method[27] respectively, $E_{Young}$ is measured to be 88.7±14.4 GPa.[21-23] In both calculation cases,[26, 27] van der Waals force was considered between $Bi_2O_2Se$ layers. In fact, the stacking of $Bi_2O_2Se$ layers is dependent on strong electrostatic interaction, it has been deduced that some unique interactions may exist between the layers.[26] This may be the reason why the experimental $E_{Young}$ is slightly higher than the theoretical value. In addition, $E_{Young}$ of few-layer $Bi_2O_2Se$ is more than 50% higher than that of 2D BP with narrow bandgap.[25] The standard deviation (16.3%) of $E_{Young}$ of $Bi_2O_2Se$ is similar with that of other few-layer 2D materials measured by the same method.[22, 25] Previous studies have demonstrated that $E_{Young}$ of bilayer 2D materials based on van der Waals stacking, such as graphene[18] and TMDCs,[21, 23] is commonly lower than that of their monolayer counterpart owing to the interlayered sliding, which dependents on the interlayered interaction.[31] This variation becomes less sensitive when the materials become thicker.[18, 25] In terms of $Bi_2O_2Se$, the strong electrostatic interaction between layers further avoids the interlayered sliding, which leads to the layer-independent property of its $E_{Young}$. It is worth mentioning that $E_{Young}$ was decreased to 42.7 GPa (Figure S9) in case of spontaneous formation of wrinkles in $Bi_2O_2Se$ flakes during the transfer procedure. Similar phenomenon in other 2D materials has been reported that wrinkled structures decrease $E_{Young}$ but improve the flexibility of graphene.[32] The above analysis suggests that the strain at hole center that applies on the $Bi_2O_2Se$ flake is isotropic and along the radial direction, as illustrated by the color-gradient arrows in the left panel of Figure 4c. The force bearing point is subjected to the largest strain (red part in the right panel of Figure 4c) and will be fractured eventually with increasing loading. To further quantitatively describe the flexibility of few-layer $Bi_2O_2Se$, radial strain at the fracture point ($\varepsilon$) is introduced by,[32]

$$\varepsilon = \frac{2(\delta_0+\delta)^2}{3r^2} \quad (5)$$

Where $\delta$ is the critical indentation before breaking. As shown in Figure 4d, for 6L Bi$_2$O$_2$Se, it can tolerate the radial strain of up to 3%, while the 40L flake undergoes fracture under 1% strain, suggesting a better mechanical compliance of thinner Bi$_2$O$_2$Se flake. Uncovering the mechanical features is favorable to enrich the database of the physical properties of 2D Bi$_2$O$_2$Se.

## 3. Conclusion

We have developed a reliable and effective PDMS-mediated method to transfer thin Bi$_2$O$_2$Se flakes from mica onto MEMS substrates with good fidelity. Taking advantage of the flexible PDMS with high adhesive force, the transferred Bi$_2$O$_2$Se, including thin ones down to 5L, maintain the initially rectangular shape, homogeneous morphology, and high-quality structures. Furthermore, the mechanical properties of 2D Bi$_2$O$_2$Se have been comprehensively studied and we disclose that the few-layer Bi$_2$O$_2$Se shows a high stiffness of 18-23 GPa and a layer-independent Young's modulus of 88.7 ± 14.4 GPa. We also show the increase of fracture radial strain with the decrease of layer numbers of Bi$_2$O$_2$Se, as well as the high flexibility of a 6L flake that withstands a deformation of more than 3%. The PDMS-mediated transfer method and experimental measurements of the mechanical properties of 2D Bi$_2$O$_2$Se pave a solid foundation for the applications of this emerging material in flexible devices.

## 4. Experimental Section

*Growth and Transfer of Few-Layer Bi$_2$O$_2$Se*: Few-layer Bi$_2$O$_2$Se samples were grown by the method of VS deposition.[8] Firstly, bulk Bi$_2$O$_2$Se and mica substrates were put into a quartz tube with the inner diameter of 1 inch, and the distance between the growth substrates and the powders is in the range of 8-12 cm. Subsequently, bulk Bi$_2$O$_2$Se were positioned at the hot center of the growth chamber (TF55035KC-1, Thermo Fisher Scientific, USA) and the mica substrates were kept downstream. Finally, the $Bi_2O_2Se$ vapor from the high-temperature zone at 700 °C was carried by the pure Ar (99.99%) with the flow rate of 200 sccm to deposit on the mica substrates for 6 min, to obtain 2D $Bi_2O_2Se$ flakes with various thickness.

For the PMMA-assisted transfer method, a PMMA layer was spin-coated on the as-grown $Bi_2O_2Se$ on the mica with the speed of 2,000 rpm for 1 min followed by the heat treatment at 150 °C for 1 min. To assist PMMA layer, commercially available PDMS with the thickness of 200 μm (KYN-200, Hangzhou Bald Advanced Materials Co., Ltd., China) was attached to the surface of PMMA/$Bi_2O_2Se$, the adhesion between which was improved by heating at 80 °C for 1 min. Then the sample (PDMS/PMMA/$Bi_2O_2Se$/mica) was put into distilled water in a petri dish to peeled off PDMS/PMMA/$Bi_2O_2Se$ from mica and transfer onto a micro-electro-mechanical system (MEMS) $SiO_2$/Si substrate with hole arrays. The diameter and depth of the holes were 2 μm and 300 nm respectively, which were confirmed by SEM and AFM (Figure S5). Afterwards, the PDMS layer was peeled off by heating the sample to 150 °C to enhance the adhesion between PMMA/$Bi_2O_2Se$ and the MEMS $SiO_2$/Si substrate but reduce their interaction with the PDMS. Finally, PMMA layer was removed by immersing the sample into pure acetone overnight to obtain $Bi_2O_2Se$ flakes onto MEMS $SiO_2$/Si substrate. However, this method was effective to transfer thick $Bi_2O_2Se$ flakes, while the thin ones were left at the initial position on mica surface.

Before transferring $Bi_2O_2Se$ flakes by the PDMS-mediated method, the as-grown $Bi_2O_2Se$ samples on mica were soaked into HF solution with an ultralow concentration of 0.01% for 5-10s to weaken the interaction between the thin flakes and mica, then cleaned in distilled water immediately. The extremely low concentration of HF solution and immediate cleaning would not deteriorate the structure of $Bi_2O_2Se$ flakes, as approved by the characterization results in this manuscript. For the PDMS-mediated method, a piece of flexible PDMS with appropriate thickness of 200 um (KYN-200, Hangzhou Bald Advanced Materials Co., Ltd., China) was attached and pressed to make the adhesion stronger between $Bi_2O_2Se$ flakes and PDMS. Afterwards, PDMS layer along with $Bi_2O_2Se$ flakes were peeled off from mica with the help of a water droplet of 5 μL, followed by transferring PDMS/$Bi_2O_2Se$ onto MEMS $SiO_2$/Si substrate. Finally, the transferred samples were heated gradually to 180 °C to improve the adhesion between $Bi_2O_2Se$ flakes and the target substrate, while bare PDMS layer was easily removed to obtain suspended $Bi_2O_2Se$ flakes with various thickness. The high adhesion and flexibility of PDMS play the most important role to successfully transfer the $Bi_2O_2Se$ flakes with various thichness, including the thin ones with the thickness of 3.2 nm (5 layers), onto the MEMS $SiO_2$/Si substrate. On the one hand, AFM contact mode was used to study that the adhesive force of PDMS is more than two times as high as that of PMMA (Figure S1), providing a stronger interaction between thin $Bi_2O_2Se$ flakes and the transfer mediate. Rather than the rigid PMMA,[2] the soft PDMS can make a more conformable contact with the $Bi_2O_2Se$ flakes, which may also help overcome the interaction between the thin flakes and mica.

*Materials Characterization*: The morphology of samples was characterized by optical microscopy (Imager A2m, Carl Zeiss, Germany) and SEM (5kV, SU8010, Hitachi, Japan). AFM (Cypher ES, Oxford Instruments, USA) was used to further characterize the morphology and measure the thickness of all samples with the tapping mode, which avoids damages during the scanning processes. Raman spectrometer (HR800, Horiba JY, Japan) was applied to collect the Raman spectra and mapping patterns. The spot diameter of the excited laser with the wavelength of 633 nm is ~ 1 μm, and the step size of Raman mappings is 1 μm. TEM characterization (Tecnai F30, FEI, USA) was performed to obtain TEM images and SAED patterns at the operating voltage of 300 kV. XPS analyses (ESCALAB 250Xi, Thermo Fisher, USA) were performed to confirm the chemical composition of $Bi_2O_2Se$ flakes.

*Measurement of Mechanical Properties of $Bi_2O_2Se$ Flakes by Nanoindentation*: After the collection of AFM images of $Bi_2O_2Se$ flakes by the tapping mode, contact mode was applied for subsequent application of force for nanoindentation measurements. The spring constant of an AFM cantilever (AC55TS, Oxford Instruments, USA) was calibrated using the thermal noisy method before each measurement. The extension and retraction speed of the AFM tip was 500 nm $s^{-1}$.

*Measurement of Adhesive Force of PDMS and PMMA*: AFM can be used to evaluate the adhensive force of polymers.[30] To prepare the PDMS sample, a piece of PDMS (KYN-200, Hangzhou Bald Advanced Materials Co., Ltd., China) with the thickness of 200 μm was directly pasted on $SiO_2$/Si substrate. And PMMA (Allresist, Germany) was spin-coated on $SiO_2$/Si substrate at a speed of 2,000 rpm for 10 min, then dried on a heating plate at 150 °C for 1.5 min to obtain the PMMA sample. Subsequently, a force of 100 nN was loaded on and unloaded from the samples by an AFM tip with the speed of 500 nN $s^{-1}$ to collect the approach and extraction force curves, respectively. Accordingly, the adhesive forces between the polymers and the AFM tip were obtained (Figure S1).

**Acknowledgement**

The authors thank the support from China Postdoctoral Science Foundation (No. 2019M650665), the National Natural Science Foundation of China (Nos. 51950410577, 51722206, 51920105002, 51991340, and 51991343), Guangdong Innovative and Entrepreneurial Research Team Program (No. 2017ZT07C341), the Bureau of Industry and Information Technology of Shenzhen for the "2017 Graphene Manufacturing Innovation Center Project" (No. 201901171523), and the Shenzhen Basic Research Project (No. JCYJ20190809180605522).

## Conflict of Interest

The authors declare no conflict of interest.

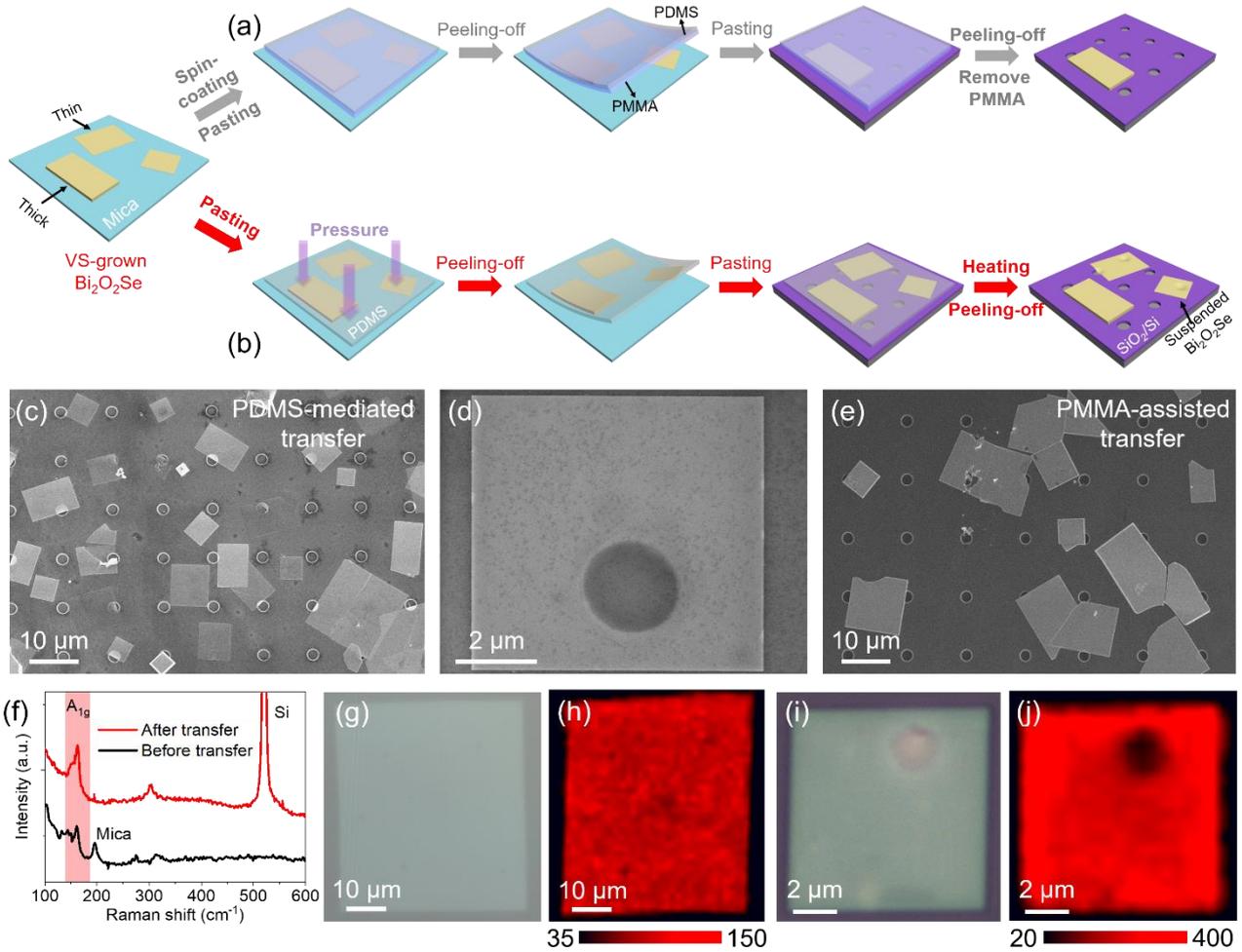

**Figure 1. Transfer of Bi$_2$O$_2$Se.** Schematic processes of transfer of Bi$_2$O$_2$Se from mica onto MEMS SiO$_2$/Si substrate by (a) PMMA-assisted and (b) PDMS-mediated methods. (c) Low-magnification and (d) high-magnification SEM images of thin Bi$_2$O$_2$Se flakes on MEMS SiO$_2$/Si substrate after the PDMS-mediated transfer. (e) SEM image of thin Bi$_2$O$_2$Se flakes on MEMS SiO$_2$/Si substrate after the PMMA-assisted transfer. (f) Raman spectra of Bi$_2$O$_2$Se before (black curve) and after (red curve) the PDMS-mediated transfer. (g) Optical microscopy image and (h) corresponding Raman A$_{1g}$ intensity map of Bi$_2$O$_2$Se on a mica substrate. (i) Optical microscopy image and (j) corresponding Raman A$_{1g}$ intensity map of Bi$_2$O$_2$Se, which was transferred onto MEMS SiO$_2$/Si substrate by PDMS-mediated method.

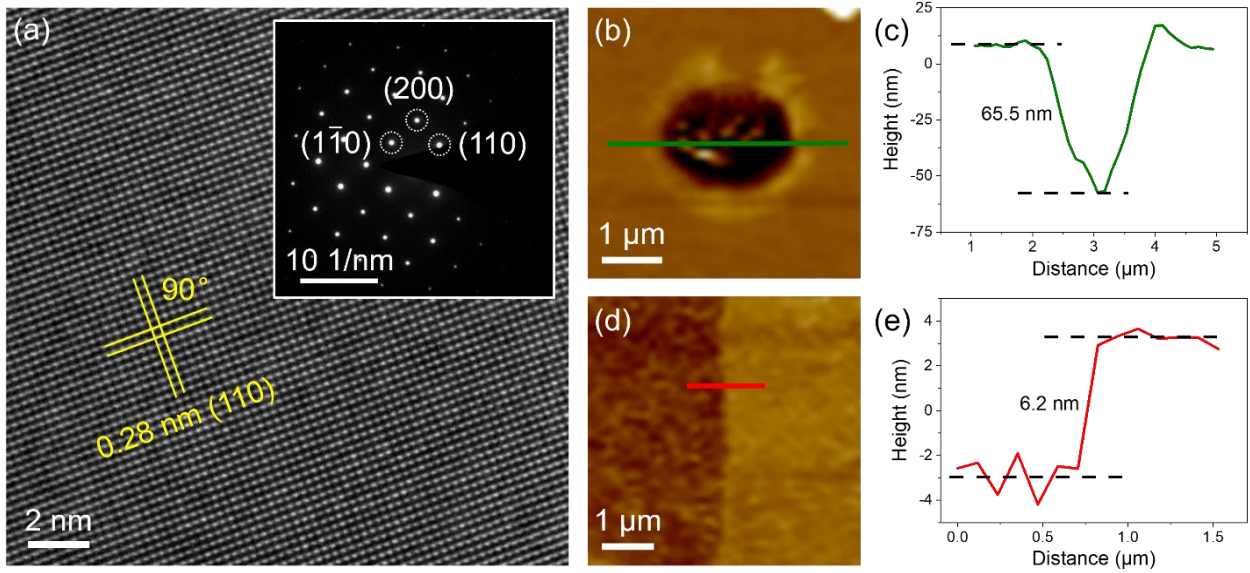

**Figure 2. Characterization of $Bi_2O_2Se$ flakes.** (a) A TEM image of $Bi_2O_2Se$. Inset is the selected-area electron diffraction pattern. (b) AFM image of a suspended $Bi_2O_2Se$ flake on MEMS $SiO_2$/Si substrate and (c) Height profile of the green line in (b). (d) AFM image of the edge of this suspended $Bi_2O_2Se$ on MEMS $SiO_2$/Si substrate and (e) Height profile of the red profile in (d), showing a thickness of 6.2 nm.

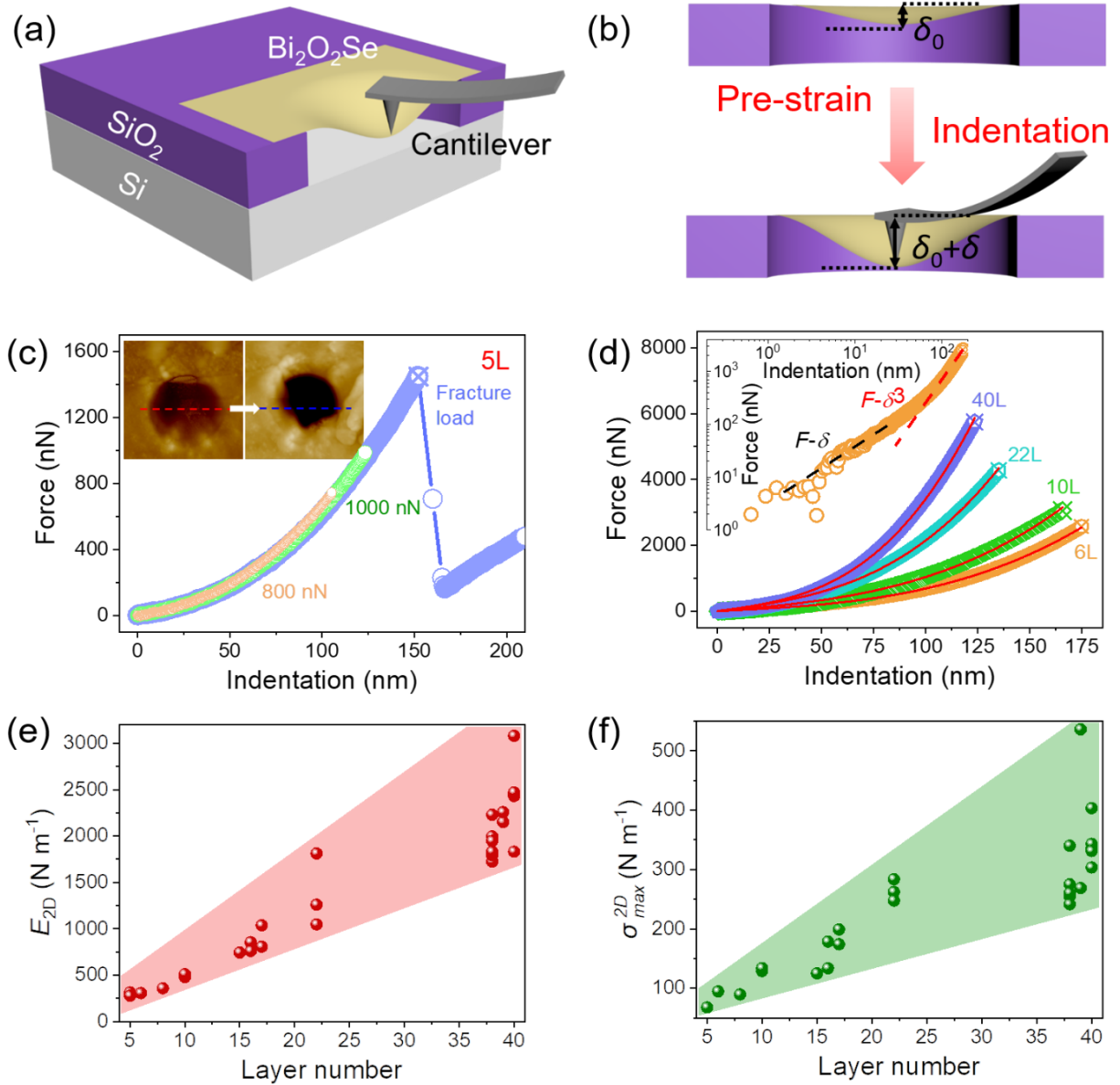

**Figure 3. Nanoindentation measurements and elastic response of $Bi_2O_2Se$ flakes with different thickness.** (a) Schematic of nanoindentation apparatus used to apply a force on suspended $Bi_2O_2Se$ flakes by an AFM tip. (b) Schematic of the deflection of $Bi_2O_2Se$ flakes under loading force. (c) $F$-$\delta$ curves of a 5L $Bi_2O_2Se$ at different loadings. Insets are AFM images of suspended $Bi_2O_2Se$ before force loading and after application of fracture load. (d) Comparisons of $F$-$\delta$ curves of $Bi_2O_2Se$ with different thicknesses at fracture load. Inset is the same curve of 6L $Bi_2O_2Se$ in logarithmic coordinates. Plots of (e) $E^{2D}$ and (f) $\sigma_{max}^{2D}$ of $Bi_2O_2Se$ flakes versus their thicknesses.

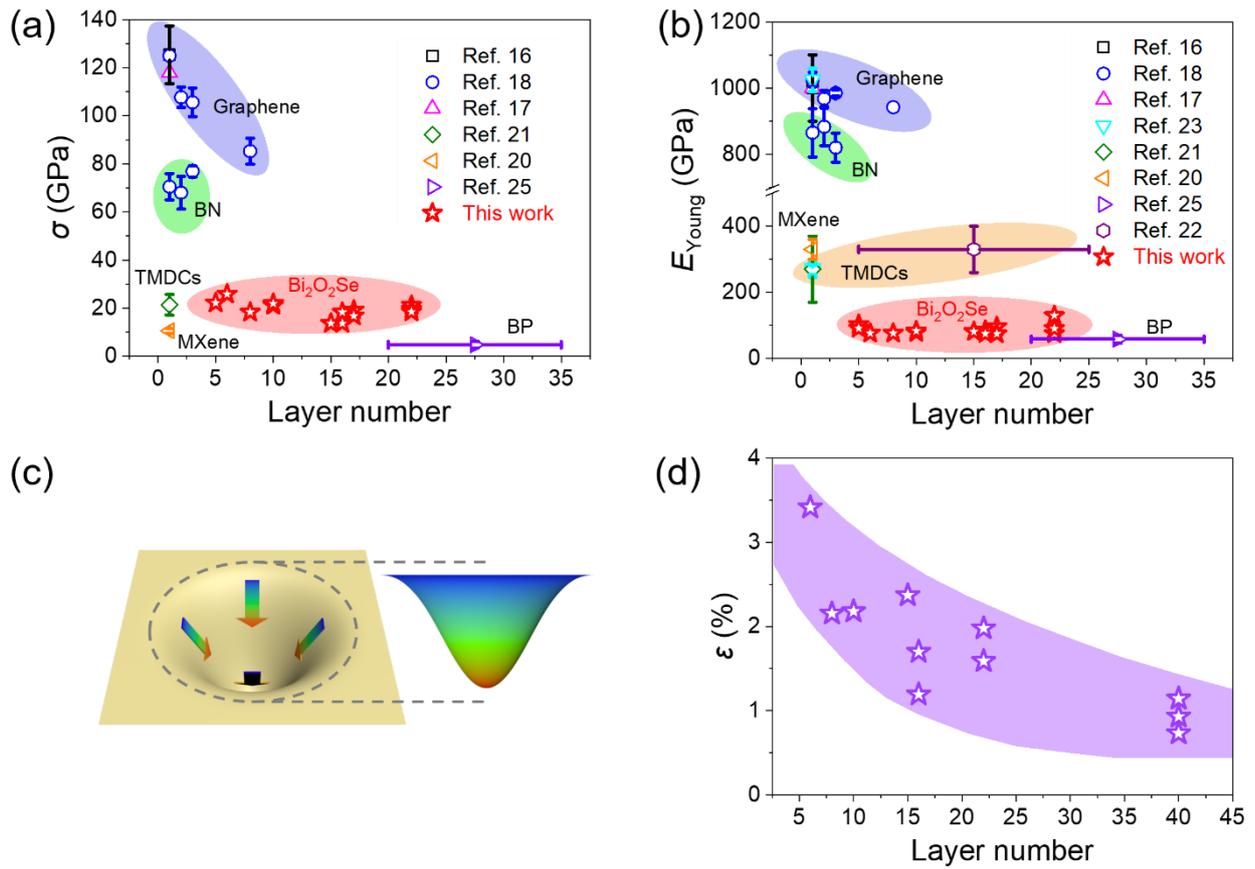

**Figure 4. Mechanical properties of Bi$_2$O$_2$Se.** Comparisons of (a) $\sigma$ and (b) $E_{\text{Young}}$ of Bi$_2$O$_2$Se and other 2D materials. (c) Schematic of isotropic deformation of Bi$_2$O$_2$Se under pressure and the radial direction is indicated by the arrays. (d) Fracture radial strain of Bi$_2$O$_2$Se with different thicknesses.